\documentclass[aps, prb, twocolumn, superscriptaddress]{revtex4-2}

\usepackage{graphicx}
\usepackage{color}

\newcommand{\gd}{KBaGd(BO$_3$)$_2$}
\newcommand{\yb}{KBaYb(BO$_3$)$_2$}
\newcommand{\mub}{$\mu_{\rm B}$}

\begin{document}
\title{Adiabatic demagnetization cooling well below the magnetic ordering temperature in the triangular antiferromagnet KBaGd(BO$_3)_2$ }

\author{A. Jesche}
  \affiliation{EP VI, Center for Electronic Correlations and Magnetism, Institute of Physics, University of Augsburg, D-86159 Augsburg, Germany}

\author{N. Winterhalter-Stocker}
  \affiliation{EP VI, Center for Electronic Correlations and Magnetism, Institute of Physics, University of Augsburg, D-86159 Augsburg, Germany} 

\author{F. Hirschberger}
  \affiliation{EP VI, Center for Electronic Correlations and Magnetism, Institute of Physics, University of Augsburg, D-86159 Augsburg, Germany} 

\author{A. Bellon}
  \affiliation{EP VI, Center for Electronic Correlations and Magnetism, Institute of Physics, University of Augsburg, D-86159 Augsburg, Germany} 

\author{S. Bachus}
  \affiliation{EP VI, Center for Electronic Correlations and Magnetism, Institute of Physics, University of Augsburg, D-86159 Augsburg, Germany} 

\author{Y. Tokiwa}
  \affiliation{EP VI, Center for Electronic Correlations and Magnetism, Institute of Physics, University of Augsburg, D-86159 Augsburg, Germany}
  \affiliation{Advanced Science Research Center, Japan Atomic Energy Agency, Tokai, Ibaraki, Japan} 

\author{A. A. Tsirlin}
    \affiliation{EP VI, Center for Electronic Correlations and Magnetism, Institute of Physics, University of Augsburg, D-86159 Augsburg, Germany} 
    \affiliation{Felix Bloch Institute for Solid-State Physics, Leipzig University, 04103 Leipzig, Germany}

\author{P. Gegenwart}
  \email[]{philipp.gegenwart@physik.uni-augsburg.de}
  \affiliation{EP VI, Center for Electronic Correlations and Magnetism, Institute of Physics, University of Augsburg, D-86159 Augsburg, Germany} 
  
\begin{abstract}
Crystal structure, thermodynamic properties, and adiabatic demagnetization refrigeration (ADR) effect in the spin-$\frac72$ triangular antiferromagnet KBaGd(BO$_3)_2$ are reported. With the average nearest-neighbor exchange coupling of 44\,mK, this compound shows magnetic order below $T_N=263$\,mK in zero field. The ADR tests reach the temperature of $T_{\min}=122$\,mK, more than twice lower than $T_N$, along with the entropy storage capacity of 192\,mJ\,K$^{-1}$\,cm$^{-3}$ and the hold time of more than 8 hours in the PPMS setup, both significantly improved compared to the spin-$\frac12$ Yb$^{3+}$ analog. We argue that KBaGd(BO$_3)_2$ shows a balanced interplay of exchange and dipolar couplings that together with structural randomness and geometrical frustration shift $T_{\min}$ to well below the ordering temperature $T_N$, therefore facilitating the cooling.

\end{abstract}

\maketitle

\section{Introduction}
Energy-efficient and sustainable cooling is one of the key demands of modern technology. 
The utilization of quantum effects, such as entanglement~\cite{spinwen2019} and superconductivity~\cite{gurevich2014}, often requires temperatures on the order of several Kelvin and even in the mK range. Gradual exhaustion of helium supplies renders the conventional cooling technique of gas compression increasingly less suitable for large-scale technological applications at low temperatures~\cite{cho2009,nuttall2012}. This calls for the development of alternative approaches to refrigeration. Adiabatic demagnetization refrigeration (ADR) is particularly suitable in this case because it entails cooling in a solid-state material by a simple sweep of the magnetic field and allows to reach sub-Kelvin temperatures~\cite{debye1926,giauque1927}. The working principle of ADR relies on the large entropy change upon applying magnetic field to a paramagnet. The lowest cooling temperature $T_{\rm min}$ is determined by residual magnetic interactions and typically coincides with the magnetic ordering temperature of the material~\cite{wikus2014}. 

In conventional ADR materials, the ultimate $T_{\min}$ values of $100-200$\,mK and below are reached by reducing magnetic interactions through separating magnetic ions with water molecules~\cite{debye1926}. However, the large amounts of crystal water contained in paramagnetic ADR salts~\cite{wikus2014} lead to their poor stability even under ambient conditions and especially upon heating. The water-based ADR salts are incompatible with high-vacuum applications that become increasingly more important in the field. 

Recently, we proposed the water-free alternative to conventional ADR salts and showed that the triangular antiferromagnet KBaYb(BO$_3)_2$ is a promising magnetic refrigerant with $T_{\min}$ below 22\,mK when feedback control of the bath is used, and 40\,mK in a PPMS setup~\cite{Tokiwa2021}. The excellent ADR performance of this material stems not only from the significant separation between the magnetic Yb$^{3+}$ ions but also from the effect of geometrical frustration on the triangular lattice and from the fact that K$^+$ and Ba$^{2+}$ randomly occupy the same position in the crystal structure~\cite{guo2019}. The random distribution of two ions with different charge leads to random crystal electric fields~\cite{li2017} and eventually causes a randomness of exchange couplings~\cite{rau2018} that lowers the magnetic ordering temperature $T_N$ or even eliminates magnetic order completely, as in the recently studied triangular spin-liquid candidates~\cite{li2020}.

The entropy storage capacity of KBaYb(BO$_3)_2$ is limited by the pseudospin-$\frac12$ nature of the Yb$^{3+}$ ion that pins the maximum available entropy at $R\ln 2$. Here, we explore low-temperature magnetism and the ADR performance of the isostructural material with spin-$\frac72$ Gd$^{3+}$ as the magnetic ion~\cite{sanders2017}. 
The use of the large local spin increases the entropy storage capacity to $R\ln 8$ and also improves the hold time. 
It simultaneously introduces magnetic order and raises $T_{\min}$, but keeps the water-free rare earth borate, KBaGd(BO$_3)_2$, far superior to water-based ADR materials with the similar $T_{\min}$. 
This becomes possible because the cooling effect in KBaGd(BO$_3)_2$ extends well below $T_N=263$\,mK. 
We discuss the possible origin of this unusual behavior.
We note that a utility model for the usage of KBa$R$(BO$_3$)$_2$ ($R$=rare earth) based pellets for UHV compatible ADR to Millikelvin temperatures has been filed in Germany by the University of Augsburg\,\cite{Utility2020}.

\section{Experimental}
Polycrystalline \gd~was grown in a two-step solid state reaction from Gd$_2$O$_3$, BaCO$_3$, K$_2$CO$_3$, and H$_3$BO$_3$.
Powders of the starting materials were thoroughly mixed, slightly pressed in an Al$_2$O$_3$ crucible and placed in a standard box furnace. 
The material was heat to 700$^\circ$C over $\sim 4$\,h and held for 24\,h followed by furnace cooling. 
The sample was taken out of the furnace when the temperature reached roughly 400$^\circ$C, which makes it easier to remove the material from the crucible (when compared to cooling to room temperature). 
Subsequently, the samples were again ground for at least 10 minutes. 
In a second step, the powder was heated to 900$^\circ$C over $\sim 4$\,h, held for 24\,h and furnace cooled to roughly 400$^\circ$C.

Sample purity was checked with lab x-ray diffraction (XRD) measurements performed with the Rigaku Miniflex diffractometer (CuK$_{\alpha}$ radiation, Bragg-Brentano geometry).
High-resolution XRD data were collected at the ID22 beamline of European Synchrotron Radiation Facility (Grenoble, France) using the wavelength of 0.35432\,\r A and the multi-analyzer detector setup. 
The powder sample was placed into a thin-wall borosilicate glass capillary and spun during the measurement. 
The data were collected at room temperature (295 K) and at 80 K using using the liquid-nitrogen cryostream. 
Crystal structure refinements were performed in the \texttt{Jana2006} program\,\cite{Petricek2014}.

Temperature- and field-dependent magnetization was measured using a Quantum Design MPMS3 magnetometer equipped with the iQuantum He3 option. 
\gd~powder with a total mass of 10.9\,mg was mounted in a plastic capsule. 

Specific heat for $T > 2$\,K was measured using a Quantum Design PPMS: a pressed pellet of 3\,mm diameter was prepared from a mixture of 7.01\,mg silver powder and 7.36\,mg of \gd~in order to improve the thermal conductivity of the material at low temperatures. 
The powder was thoroughly mixed and pressed at 210\,MPa. 
The mass of the obtained pellet, however, was found to be $m = 13.59$\,mg, somewhat smaller than the sum of the starting materials.  
Roughly 5\% of the material remained in the pressing dyes.
However, the composition of lost material was found to be similar to the nominal one.
The low temperature specific heat ($T < 2$\,K) was measured on a 2.73 mg piece of the \gd-silver pellet. 
The measurement was performed with a home-built setup installed in a dilution refrigerator equipped with a 13 T magnet. 
The relaxation method was used to evaluate the total heat capacity of the pellet and the silver contribution to the heat capacity was subtracted using reference measurements as explained in Sec. III.C.
Cooling by adiabatic magnetization was investigated using a home-made setup mounted on a standard 'Quantum Design' PPMS puck\,\cite{Tokiwa2021}.
A length of a plastic straw was used to hold the pellet roughly 35\,mm above the puck.
The pellet used had a total mass of 3.51\,g, a diameter of 15.0\,mm and a thickness of 5.0\,mm. 
It was pressed at 10\,kN and consisted of 50\,weight-\% \gd~and 50\,weight-\% Ag powder. 
A RuO$_2$ thermometer was glued to the sample and the resistivity measured with an excitation current of 1\,nA leading to a maximum resistivity of 13.5\,k$\Omega$ at lowest temperature. 
The measurement was performed using the four-point technique at a frequency of 13.7\,Hz with a Model 372 AC Resistance Bridge from Lake Shore Cryotechnics which uses an internal lock-in amplifier. This helped to reduce self heating effects caused by the temperature measurement. The data was read by a self written Lab View program.

\section{Results}

\subsection{Structural characterization}

\begin{figure}
 \includegraphics[width=0.47\textwidth]{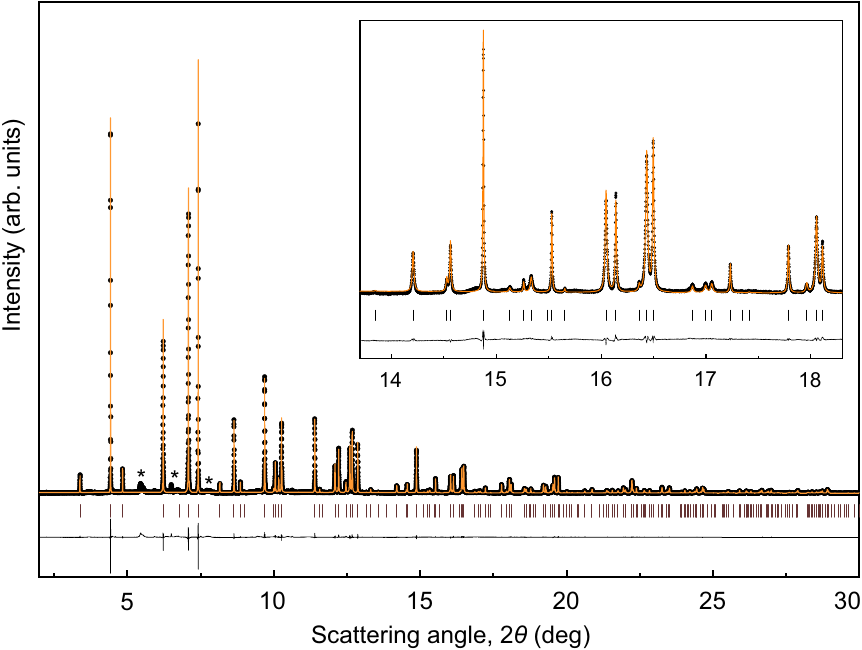}
 \caption{Rietveld refinement of the room-temperature high-resolution XRD data. Ticks show reflection positions for the $R\bar 3m$ space group, the line in the bottom is the difference pattern. The asterisks mark the impurity peaks from Gd$_2$O$_3$.
 \label{fig:xrd} 
 }
\end{figure}

Previous studies have revealed that triangular rare-earth borates may show both ordered and disordered arrangements of the alkaline and alkaline-earth ions~\cite{guo2019b}, as well as weak monoclinic distortions~\cite{guo2019c}. 
Therefore, we used high-resolution powder XRD to confirm that our sample has the same rhombohedral symmetry as the \yb~ADR material studied by us earlier\,\cite{Tokiwa2021}.
High-resolution powder XRD data (Fig.~\ref{fig:xrd}) confirm rhombohedral crystal structure of KBaGd(BO$_3)_2$ with the mixed cation site randomly occupied by K and Ba. Several weak reflections belong to the unreacted Gd$_2$O$_3$ and to an unknown phase, presumably a mixed K-Ba borate. 
The slightly anisotropic reflection widths were refined using two Lorentzian parameters, LY and LYe, that contribute to the full-width-at-half-maximum (FWHM) as

\begin{equation}
 {\rm FWHM}=({\rm LY}+{\rm LYe}\cos\varphi){\rm tan}\theta,
\end{equation}

with ${\rm LY}=7.5(1)\times 10^{-2}$\,deg and ${\rm LYe}=1.01(1)\times 10^{-1}$\,deg, whereas $\varphi$ is the angle between the diffraction vector and the broadening direction (001). 
This reflection broadening LYe indicates that the crystal structure is more coherent in the $ab$ plane where GdO$_6$ octahedra form a triangular network, while featuring defects, probably stacking faults along $c$. 
Using the Scherrer formula for the mean crystallite size~\cite{Langford1978}, $d=\lambda/(\beta\cos\theta)$ where $\beta=(\pi/2){\rm FWHM}$ is the integral breadth of the reflection corrected for the instrumental broadening obtained by measuring the Si standard (${\rm LY}_{\rm Si}=1.4\times 10^{-2}$ deg), we estimate that $d$ is of the order of 200 - 300 nm in our sample. 

Table~\ref{tab:coords} lists atomic positions and displacement parameters obtained by the Rietveld refinement. 
No significant changes are observed on cooling, apart from the reduction in the displacement parameters caused by the freezing of lattice vibrations at 80\,K.

\begin{table}
\caption{\label{tab:coords}
Atomic positions and atomic displacement parameters $U_{\rm iso}$ (in $10^{-2}$\,\r A$^2$) for KBaGd(BO$_3)_2$ at 80\,K (upper row) and 295\,K (lower row). The space group is $R\bar 3m$. The lattice parameters are $a=5.46680(1)$\,\r A and $c=17.9277(1)$\,\r A at 80\,K and $a=5.47409(1)$\,\r A and $c=17.9837(1)$\,\r A at 295\,K. 
}

\begin{ruledtabular}
\begin{tabular}{cccccc}
 Atom &      & $x/a$ & $y/b$ & $z/c$ & $U_{\rm iso}$ \\\hline
 K/Ba & $6c$ & 0 & 0 & 0.78798(5) & 0.77(2) \\
      &      & 0 & 0 & 0.78735(5) & 1.28(3) \\
 Gd   & $3a$ & 0 & 0 & 0 & 0.27(1) \\
      &      & 0 & 0 & 0 & 0.64(2) \\
 B    & $6c$ & 0 & 0 & 0.4136(5) & 0.5(2) \\
      &      & 0 & 0 & 0.4109(5) & 1.2(2) \\
 O    & $18h$ & 0.5203(3) & 0.4797(3) & 0.7472(2) & 0.50(7) \\
      &       & 0.5211(3) & 0.4789(3) & 0.7469(2) & 0.78(7) \\
\end{tabular}
\end{ruledtabular}
\end{table}

\subsection{Magnetization}
\begin{figure}
 \includegraphics[width=0.47\textwidth]{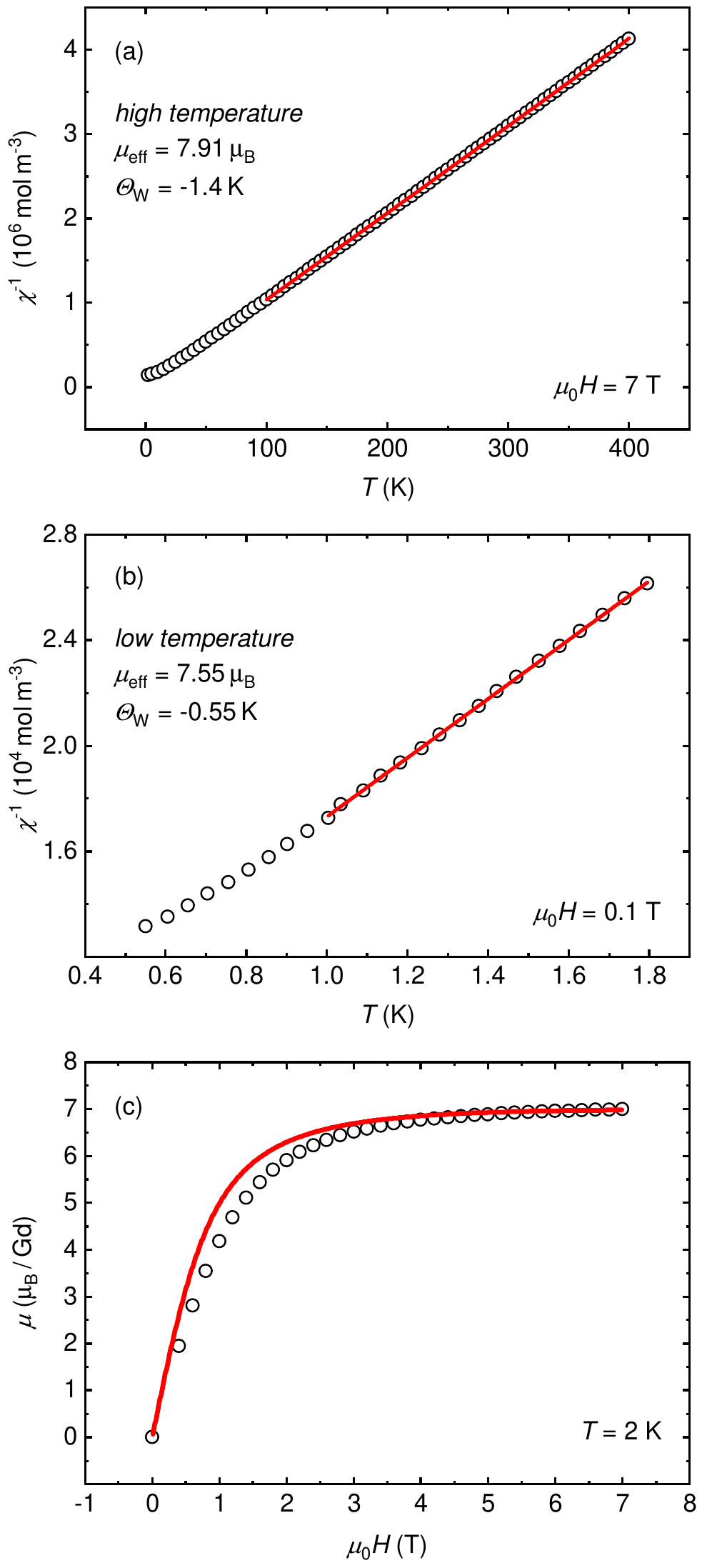}
 \caption{Magnetic properties of \gd. (a) Curie-Weiss behavior with an effective moment very close to the one of free Gd$^{3+}$ (7.94\,$\mu_{\rm B}$) is observed over a wide temperature range. (b) Minor deviations from Curie-Weiss behavior are found for $T < 1$\,K. 
( The red lines show fits to the experimental data.)
 (c) The isothermal magnetization approaches the expected saturation value of 7\,$\mu_{\rm B}$.
 The red line shows the theoretical curve for free Gd$^{3+}$ based on the Brillouin function. 
\label{mag} 
 }
 \end{figure}
Basic magnetic properties of \gd~are presented in Fig.\,\ref{mag}.
In the high-temperature region $T = 100 - 400$\,K, the magnetic susceptibility $\chi = M/H$ shows Curie-Weiss behavior with an effective magnetic moment of $\mu_{\rm eff} = 7.91$\,\mub, close to the value of 7.94\,\mub~expected for the ground state of free Gd$^{3+}$ with $S = 7/2$ (Fig.\,\ref{mag}a).
The Weiss temperature of $\Theta_{\rm W} = -1.4$\,K indicates weak antiferromagnetic interactions.
A temperature-independent contribution was considered in the fitting (red, solid line) and was found to account for less than 1\,\% of the room temperature value of $\chi$. 

Fig.\,\ref{mag}b shows $\chi^{-1}$ in the low-temperature region $T < 2$\,K.
The effective magnetic moment of $\mu_{\rm eff} = 7.55$\,\mub~is somewhat reduced compared to its high-temperature value.
Below $T = 1$\,K, the departure of $\chi^{-1}$ from the linear behavior may be caused by an onset of antiferromagnetic spin correlations or by an effect of the applied field that breaks the linear relation between $M$ and $H$ at such low temperatures.
The low-temperature Weiss temperature is determined to $\Theta_{\rm W} = -0.55$\,K.
No temperature-independent contribution was considered since the local magnetic moment of Gd$^{3+}$ is assumed to dominate over all other possible sources at such low temperatures.

The isothermal magnetization at $T = 2$\,K is plotted in Fig.\,\ref{mag}c together with the calculated one. 
The smaller slope at low fields is in accordance with the negative Weiss temperature and the proposed antiferromagnetic (AFM) ordering (see below).

\subsection{Specific heat}

\begin{figure}
 \includegraphics[width=0.47\textwidth]{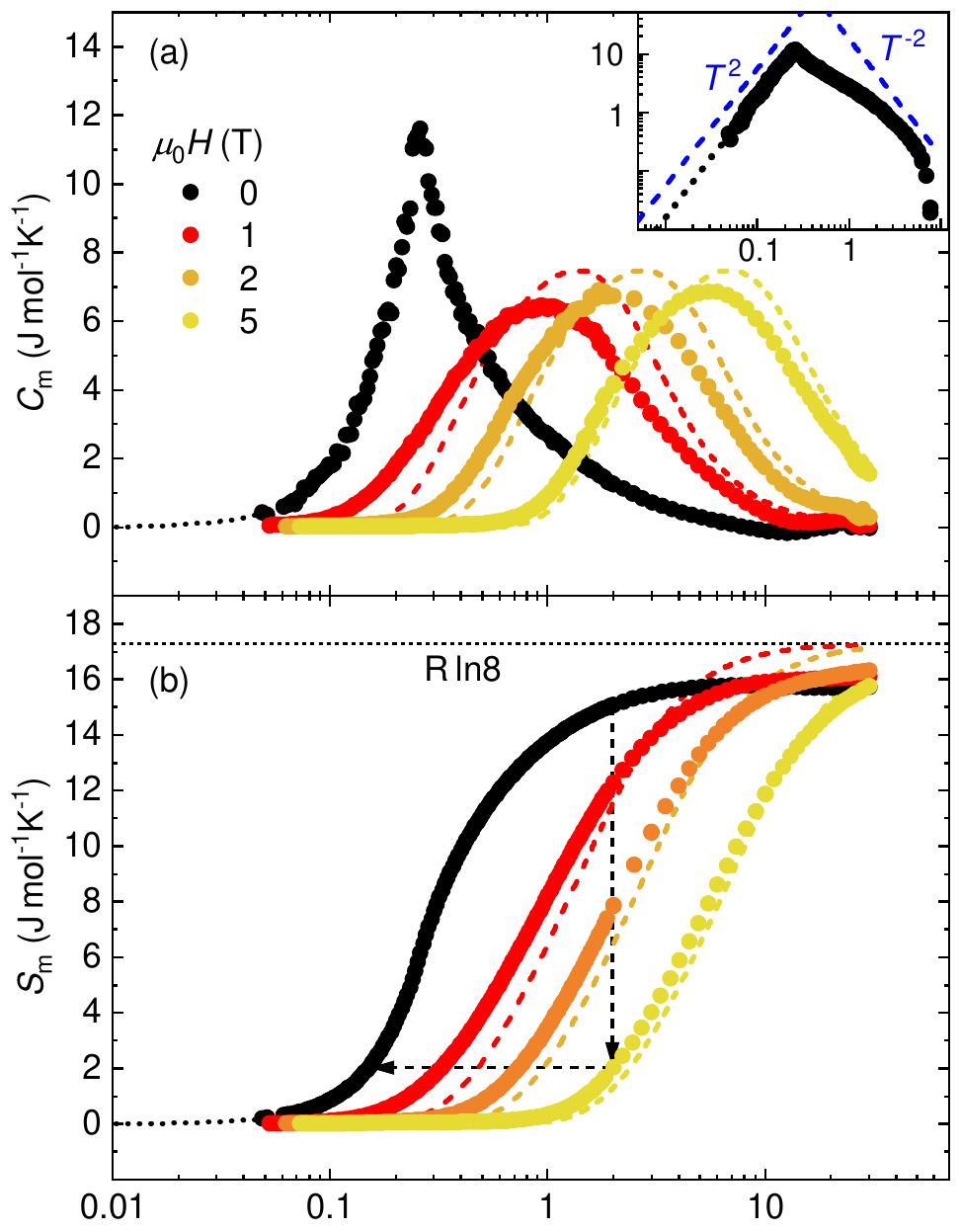}
 \caption{Magnetic contribution to the specific heat of \gd. Dashed lines show theoretical values calculated for the $S = 7/2$ multiplet of Gd$^{3+}$.
 (a) A well-defined peak centered at $T_{\rm N} = 263$\,mK marks the onset of AFM ordering for $H = 0$.
 The dotted line denotes extrapolated values (see text).
 In larger applied fields (values given in the plot), a broad Schottky-type anomaly forms.
 The inset shows the zero-field data on a double-logarithmic scale. 
 (b) The magnetic entropy approaches values of up to $\sim 94$\,\% of $R\,{\rm ln}8 = 17.3$\,J\,mol$^{-1}$K$^{-1}$ at $T = 30$\,K. 
 The entropy changes employed for cooling by adiabatic demagnetization are indicated by arrows. 
 }
 \label{hc}
\end{figure}

Figure\,\ref{hc}a shows the magnetic contribution to the specific heat $C_{\rm m}$ of \gd.
The silver contribution was subtracted based on a similar measurement on a silver pellet.
The phonon contribution was estimated by fitting the zero-field data for $T = 15-30$\,K assuming a temperature dependence proportional to $T^3$ and $T^5$ (the latter in order to account for deviations from Debye behavior at elevated temperatures). 
The obtained phonon contribution is considered field-independent and was subtracted from the experimental data. 

Magnetic ordering is inferred from a well-defined peak centered at $T_{\rm N} = 263$\,mK where the antiferromagnetic nature of the transition is evident from the field-dependence in small applied fields $\mu_0H < 0.1$\,T (see below) and the negative Weiss temperature.
The values for $T < 50$\,mK, plotted as dotted lines in Fig.\,\ref{hc}, are extrapolated from the experimental data in the temperature range $50$\,mK $<T<120$\,mK by assuming $C \sim T^\alpha$ with $\alpha = 2.04$.
This has been done in order to estimate the entropy at lowest temperatures.

In an applied field of $\mu_0H = 1$\,T the peak shape in $C(T)$ changes to a Schottky-like anomaly with a maximum at $T = 1.0$\,K.
The maximum is shifted to higher temperatures in larger applied fields. 
The dashed lines show theoretical curves for non-interacting Gd moments at the same values of the magnetic field (calculated from the temperature derivative of $E_{\rm mag} = -\mu_{\rm sat} B_J B$, $B_J$:= Brillouin function). These curves are systematically shifted to higher temperatures, indicating antiferromagnetic interactions between the Gd$^{3+}$ spins.
The zero-field data for $T < T_{\rm N}$ nicely follows a $T^2$ behavior as can be seen from the double-logarithmic plot in the inset of Fig.\,\ref{hc}a.
We find an exponent of $\alpha = -1.0$ between $T_{\rm N}$ and $T = 1$\,K.
Above $T = 3$\,K, $C_m$ approaches the $1/T^2$ regime expected in the high-temperature limit\,\cite{Johnston2000}. 
However, contributions from silver and phonons become dominant in this temperature range and prevent us from a further analysis of the high-temperature limit.

Fig.\,\ref{hc}b displays the magnetic entropy $S_{\rm m}$ obtained by integrating $C_{\rm m}/T$. 
For $H = 0$, the contribution of the extrapolated low temperature data amounts to $0.2$\,J\,mol$^{-1}$K$^{-1}$.
The entropy becomes temperature-independent for $T \geq 7$\,K and levels in at $S_{\rm m} \approx 15.7$\,J\,mol$^{-1}$K$^{-1}$ that corresponds to 91\% of $R\,{\rm ln}8$, which is expected for the fully degenerate $S = 7/2$ ground state multiplet of Gd$^{3+}$.
Higher values of $S_{\rm m}$ are found for larger applied fields: 94\% of $R\,{\rm ln}8$ in $\mu_0H = 2$\,T. 
Furthermore, there is significant increase with temperature even around $T = 30$\,K in stark contrast to the zero-field data.
This renders the inaccuracy in the ratio of \gd~and silver or the phonon subtractions unlikely as a source for the missing entropy and indicates a possible residual entropy for $T \rightarrow 0$ in $H = 0$.

The arrows in Fig\,\ref{hc}b roughly depict the magnetization process employed for cooling (see below). The magnetic entropy is reduced by 13.1\,J\,mol$^{-1}$K$^{-1}$ when applying a field of $\mu_0H = 5$\,T at $T = 2$\,K (isothermal). 
Driving the field back to $H = 0$ (adiabatically), is supposed to cool the material to below $T = 150$\,mK.

\begin{figure}
 \includegraphics[width=0.47\textwidth]{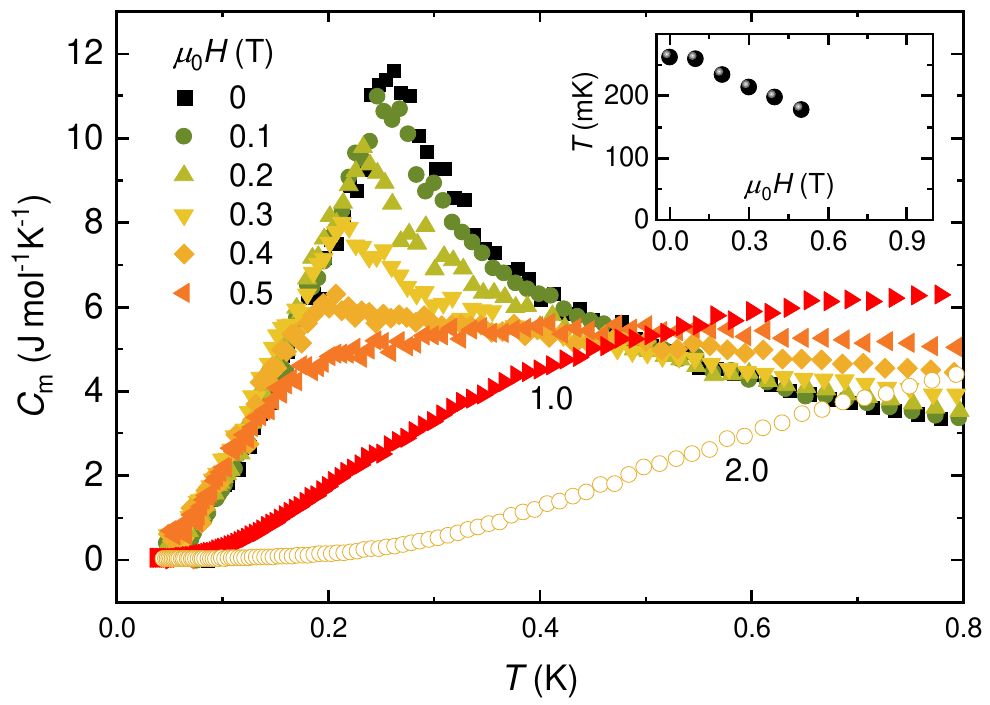}
 \caption{Specific heat of \gd~in the vicinity of the AFM ordering. 
 Significant field-dependence in the magnetically ordered state is observed down to temperatures as low as half of $T_{\rm N} = 263$\,mK.
 The inset shows the field dependence of the peak in $C(T)$. 
 }
 \label{hc-lt}
\end{figure}

Figure\,\ref{hc-lt} shows the low-temperature specific heat in various applied fields for temperatures below $T = 0.8$\,K. 
For $\mu_0H \leq 0.4$\,T, a clear peak is observable, which indicates magnetic ordering and shifts to lower temperature with increasing applied field as expected for an AFM transition.
Obviously, specific heat and corresponding entropy are strongly field-dependent for temperatures down to $T \approx 125$\,mK that is less than half of $T_{\rm N}$.
The field-dependence of the peak in $C(T)$ is depicted in the inset of Fig.\ref{hc-lt}. 
At lower temperature, however, $C_{\rm m}$ is basically field-independent for $T < 100$\,mK and $\mu_0H < 0.5$\,T.
In fields larger than $\mu_0H \approx 0.5$\,T, a crossover to a Schottky anomaly takes place and the AFM transition is fully suppressed at $\mu_0H = 1.0$\,T.

\subsection{Cooling by adiabatic demagnetization}

\begin{figure}
 \includegraphics[width=0.47\textwidth]{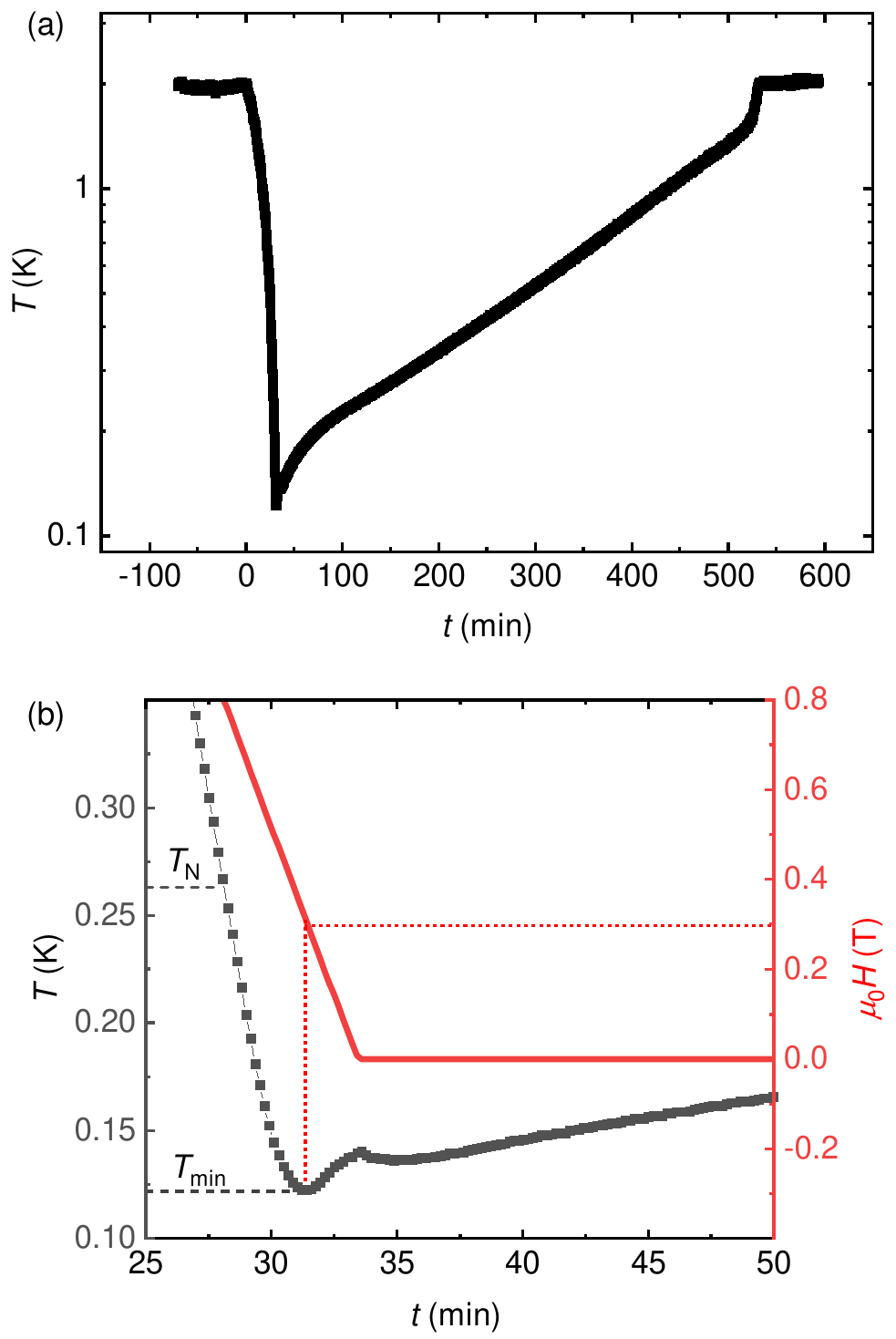}
 \caption{Cooling of \gd~by adiabatic demagnetization. (a) Beginning from $\mu_0H = 5$\,T at $t = 0$, the applied field is ramped to zero over 33 minutes (2.5\,mT per second). The temperature decreases to a minimum of $T_{\rm min} = 122$\,mK. 
 After reaching $H = 0$, the sample slowly warms up due to finite thermal coupling and reaches the sample chamber temperature of $T = 2$\,K after 8 hours and 19 minutes. 
 (b) $T_{\rm min}$ is reached at an applied field of $\mu_0H = 0.3$\,T and a local maximum is observed for $H = 0$.
 }
 \label{adr}
\end{figure}

We will now test the refrigeration performance of our material using the practical ADR cooling setup reported in Ref.\,\onlinecite{Tokiwa2021}.
The sample was cooled to $T = 2$\,K in a field of $\mu_0H = 5$\,T and the 'high vacuum mode' of the PPMS ($p < 10^{-4}$\,mbar) was employed in order to achieve thermal decoupling.
Then the applied field is ramped to $H = 0$ at a rate of $\mu_0H = 2.5$\,mT per second.
An instantaneous decrease of the sample temperature is observed as soon as the field starts to change, which defines $t = 0$.
Fig.\,\ref{adr}a shows the measured temperature evolution over a time frame of more than 10 hours. 
After reaching the minimum temperature of $T_{\rm min} = 122$\,mK, it takes more than 8 hours until the sample warms up to the chamber temperature that is constantly held at $T = 2$\,K. 
Note that thermal transport takes place via the manganin wires, the thin plastic straw that holds the sample, and a finite amount of exchange gas.

On closer inspection, it can be seen that the minimum temperature is already reached before the external field is driven to zero (Fig.\,\ref{adr}b); at $T_{\rm min}$ we find $\mu_0H = 0.3$\,T. 
When the applied field reaches $H = 0$, however, the temperature slightly decreases again.

\begin{figure}
 \includegraphics[width=0.47\textwidth]{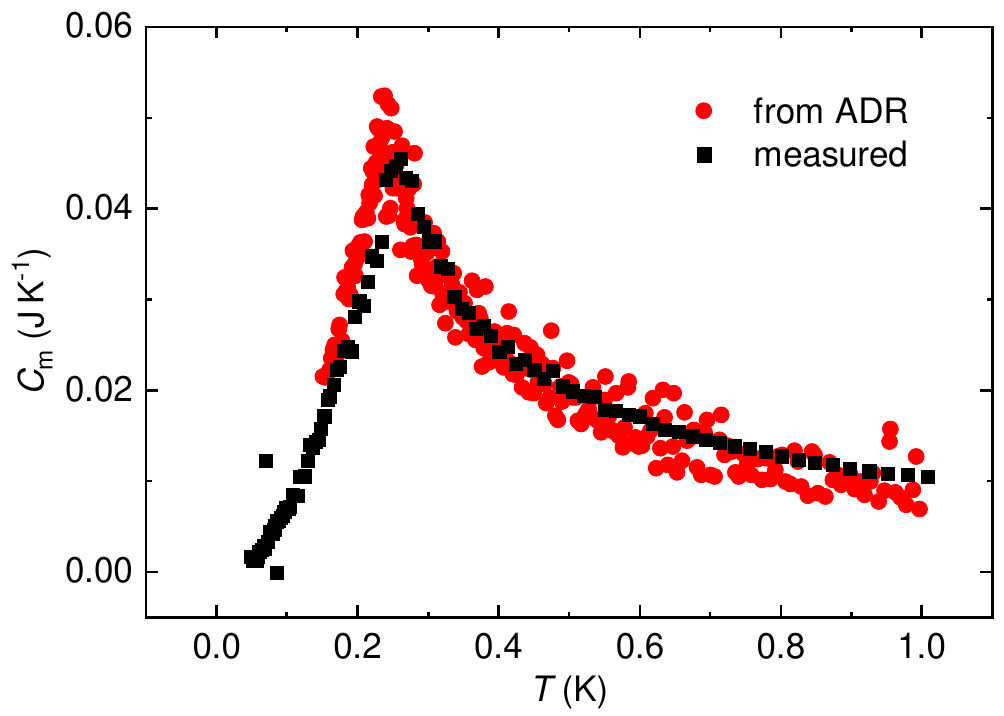}
 \caption{Total (magnetic) heat capacity of the ADR pellet derived from the warming curve of the ADR experiment, $C_{\rm ADR}$, and calculated from the measured molar heat capacity. 
The former was obtained by assuming $\dot{Q} = C_{\rm ADR} \dot{T}$ with a constant heat input of $\dot{Q} = 42.5$\,$\mu$W. 
 }
 \label{hcadr}
\end{figure}

There is no sharp anomaly visible at $T_{\rm N}$, neither during cooling in decreasing field nor during warming in zero field. 
However, the derivative of $T$ with respect to $t$ during warming reveals a corresponding signature that is directly related to the specific heat (Fig.\,\ref{hcadr}).
It is assumed that the following equation holds true for the warming part of $T(t)$ in zero field
\begin{equation}
\dot{Q} = C_{\rm ADR} \dot{T},
\end{equation}
with $\dot{Q}$ being a constant heat input (per time), $C_{\rm ADR}$ the total heat capacity of the ADR pellet, and $\dot{T}$ the time-derivative of the temperature in the ADR experiment (see Fig.\,\ref{adr}).
In order to obtain a smooth derivative, $\dot{T}$, regions of 160 time and temperature data points were averaged. (The $C_{\rm ADR}$ data shown in Fig.\,\ref{hcadr} is based on a 6.5 hour measurement with two data points recorded per second.) 
Good agreement between the 'directly' measured heat capacity (see previous section) and $C_{\rm ADR}$ is obtained for choosing $\dot{Q} = 42.5$\,$\mu$W.
It is noticeable that $C_{\rm ADR}$ is somewhat larger than the directly obtained values for $T < T_{\rm T}$ whereas the opposite is observed at higher temperatures. 
This likely reflects a finite temperature dependence of $\dot{Q}$ that has been neglected. 
Note that assuming a linear increase of $\dot{Q}$ with temperature does not lead to an acceptable agreement with the measured data. 
A heat input of $\dot{Q} = 42.5$\,$\mu$W was confirmed on \yb~using the same ADR setup with identical parameters and comparing with the measured specific heat data.

Before turning to the discussion of our results, we want to elaborate on the heat transfer during the ADR experiment, that is the deviation from adiabatic conditions. 
Important characteristics, in particular the minimum temperature achieved and the hold time (before warming back to $T = 2$\,K), are not purely intrinsic properties of the material but also influenced by the experimental setup. 
Heat transfer to the sample takes place during the whole ADR experiment and is primarily given by three contributions: a) convection by residual He gas, b) Joule heating of the thermometer, c) and thermal conduction via the sample holder (straw) and measurement wires. There are further possible contributions by eddy current heating during the field sweeps. (Note that the results provided in the following sections a)-c) were obtained on \yb.) 

Regarding a): The residual He gas is a major contributor; first experiments with a similar ADR setup do show a doubling of the hold time when performing the evacuation of the sample chamber at $T = 4$\,K instead of $T = 2$\,K - and thus creating a better vacuum.

Regarding b): Heat dissipation by the thermometer current is likely negligible since the corresponding Joule heating of $P = U \cdot I \sim 10^{-14}$\,W is orders of magnitude smaller than the estimated heat input of $\dot{Q} = 42.5$\,$\mu$W.
Nevertheless, higher excitation currents of $I \sim 100$\,nA do lead to significant heating effects at $T \sim 100$\,mK.
However, in this case presumably a local heating of the thermometer takes place that compromises the temperature reading instead of warming up the whole sample.  

Regarding c):
The situation is less clear for the thermal conduction: 
Replacing the straw by a more stable, 3d-printed sample holder (made from Formlabs 'Grey Resin V4') of similar cross-section reduces the hold time to roughly 1/3 of the initial one. 
When only the adapter that connect to the upper and lower part of plastic straw was replaced by a 3d-printed part, the hold time increases by roughly 10\%.  
This behavior indicates a significant contribution to the heat transfer that is hard to quantify any further. 
The same holds true for the wires that connect the thermometer; using insulated superconducting wires instead of ones made from manganin leads to a slight decrease in the hold time by 10\%, which indicates that the conductivity of the non-superconducting matrix (CuNi) outweighs the lower thermal conductivity of the superconducting core.

\section{Discussion and Summary} 

In stark contrast to the iso-structural \yb\,\cite{Tokiwa2021}, a clear transition into a magnetically ordered state is found for \gd.
This is not fully unexpected given that Gd$^{3+}$ carries a significantly larger magnetic moment than the CEF ground state doublet of Yb$^{3+}$ in \yb.
Assuming that dipolar and exchange interaction are proportional to the square of the magnetic moment and considering the (close to) saturation values (see Tab.\,\ref{tab:mag}), we estimate the ordering temperatures to differ by a factor of $(\mu_{\rm sat}^{\rm Gd}/\mu_{\rm sat}^{\rm Yb})^2 =  (7.0/1.3)^2 = 29$.
Thus a value of $T_N = 9$\,mK is obtained for \yb~that is in good agreement with the estimated 8\,mK based on the Weiss temperature and the assumption of weakly coupled triangular planes\,\cite{Tokiwa2021}. 
In contrast, there seems to be no direct relation between the lowest ADR temperature and hold time with other magnetic properties when comparing \gd~and \yb, see Tab.\,\ref{tab:mag}.

\begin{table}
\caption{\label{tab:mag}
Magnetic and thermodynamic properties of KBa$R$(BO$_3$)$_2$ with $R$ = Yb and Gd. Values given refer to the crystal electric field ground state multiplet. 
}

\begin{ruledtabular}
\begin{tabular}{clc}
 	  							& $R$=Yb   		& 	$R$=Gd 	\\\hline
  $S_{\rm eff}$ 					& 1/2			&	7/2		\\
$\mu_{\rm eff} (\mu_{\rm B})$    & 2.28	    		&	7.55		\\
$\mu_{\rm sat} (\mu_{\rm B})$   	& 1.4$^*$		& 	6.98 	\\
$\Theta_{\rm W}$ (mK)			& -60			&	-550		\\
$T_{\rm N}$ (mK)					& 8$^*$			&	263		\\
lowest ADR temperature (mK)$^{**}$	& 40		&	122		\\
ADR hold time (minutes)$^{**}$		& 40				&	496		\\
\end{tabular}
\end{ruledtabular}
$^*$ estimated, 
$^{**}$ measured in the PPMS
\end{table}

When comparing the heat capacity of \gd~with the actual ADR experiment, a consistent picture emerges:
There is no longer a sizable field dependence of $C(T)$ for cooling below $T \approx 125$\,mK (see smaller fields in Fig.\,\ref{hc}), which is very close to the lowest ADR temperature of 122\,mK.
In turn, this lowest temperature is obtained at a field of $\mu_0H = 0.3$\,T, close to the terminal field for the magnetically ordered state as inferred from the vanishing phase transition anomaly in $C(T)$.

We now turn to the microscopic aspects of the KBaGd(BO$_3)_2$ magnetism. 
Using the mean-field expression for a spin-$\frac72$ triangular antiferromagnet, $\Theta_W=6J\cdot S(S+1)/3=31.5J$, we estimate the nearest-neighbor exchange coupling of $J=44$\,mK. 
This value is probably an average because the random distribution of differently charged ions, in this case K$^+$ and Ba$^{2+}$, typically leads to a randomness of exchange couplings as recently proposed for YbMgGaO$_4$ where this randomness can be the main cause of the putative spin-liquid behavior\,\cite{li2020,Zhu2017}. 
An indirect indication for this randomness is the shape of the specific-heat anomaly at $T_N$. 
Contrary to the $\lambda$-type peak observed in other large-$S$ triangular antiferromagnets~\cite{svistov2006,smirnov2009}, the transition anomaly in KBaGd(BO$_3)_2$ has an almost triangular shape that may be expected in a system with a distribution of exchange couplings. 
Similar triangular-shaped anomalies have been seen in magnets with complex mechanisms of frustration due to multiple competing exchange couplings~\cite{lebernegg2016}. 
Therefore, another potentially important effect is the interlayer frustration caused by the lateral shift of adjacent triangular layers in the rhombohedral crystal structure~\cite{guo2019}. 
A similar situation has been reported in ACrO$_2$ (A = H, Na, K) where it leads to a broad regime with 2D fluctuations that precedes the onset of long-range magnet order\,\cite{somesh2021}. 
HCrO$_2$ also does show a $T^2$-dependence of $C_m$ for $T < T_{\rm N}$\,\cite{somesh2021} as observed for \gd~(see inset of Fig.\,\ref{hc}a).
Both effects -- randomness and interlayer frustration -- should lead to a broadening of the specific-heat anomaly and a shift of entropy toward lower temperatures, thus extending the ADR effect to temperatures below $T_N$.

It is further worth noting that the nearest-neighbor exchange coupling of 44\,mK in KBaGd(BO$_3)_2$ is about three times larger than the dipolar coupling of 15\,mK (dipole-dipole interaction energy of 186\,mK assuming $g=2$ and $S=\frac72$) expected between the Gd$^{3+}$ spins-$\frac72$ in this crystal structure. By contrast, Yb$^{3+}$ borates with similar metal-metal distances of about 5.5\,\r A are usually dominated by dipolar couplings~\cite{guo2019b,bag2021,jiang2022}. Therefore, the larger magnetic moment of Gd$^{3+}$ triggers an exchange component of the magnetic interaction, in addition to the dipolar one. An interesting observation here is that the nearest-neighbor exchange interaction on the triangular lattice should cause $120^{\circ}$ magnetic order~\cite{huse1988}, whereas a triangular magnet with purely dipolar couplings does not order down to zero temperature~\cite{keles2018}. This dissimilar effect of the two components of the magnetic interaction may be yet another ingredient of the unusual transition anomaly in KBaGd(BO$_3)_2$. 

Finally, we compare KBaGd(BO$_3)_2$ to the water-based ADR salts. 
The $T_{\min}$ of 122\,mK and the maximum entropy storage capacity of $S_{\rm GS}=192$\,mJ\,K$^{-1}$\,cm$^{-3}$ are superior to those of Mn(NH$_4)_2$(SO$_4)_2\cdot 6$H$_2$O (MAS) with $T_{\min}$ of 170\,mK and $S_{\rm GS}=70$\,mJ\,K$^{-1}$\,cm$^{-3}$~\cite{vilches1966}, and render our material a good choice for the first cooling stage before a material with $T_{\min}<50$\,mK, such as KBaYb(BO$_3)_2$~\cite{Tokiwa2021}, is used. 
It also outperforms many of the water-free ADR materialsn\,\cite{jang2015,Shimura2022,Kleinhans2022}.
The capability of cooling to temperatures well below $T_N$ appears to be a crucial ingredient for designing an ADR material with high entropy storage capacity and low $T_{\min}$.
\\
\newline
Note added: during the review of our manuscript, a concurrent report on \gd~has appeared\,\cite{Xiang2023-arxiv}. 
Using Monte-Carlo simulations of the magnetic susceptibility, Xiang {\textit et al.} confirm that dipolar couplings are similar in magnitude to the nearest-neighbor exchange. 
Their magnetization and specific heat data, as well as the cooling performance, are similar to our results. 

\section*{Acknowledgments}
We thank Alexander Herrnberger and Klaus Wiedenmann for technical support.
Andreas Honecker is gratefully acknowledged for useful discussions and for organizing an international research network on strongly correlated electron systems as advanced magnetocaloric materials. 
We also acknowledge ESRF for providing the beamtime and thank Andy Fitch, Catherine Dejoie, and Ola Grendal for their technical support during the measurements.


\begin{thebibliography}{33}%
\makeatletter
\providecommand \@ifxundefined [1]{%
 \@ifx{#1\undefined}
}%
\providecommand \@ifnum [1]{%
 \ifnum #1\expandafter \@firstoftwo
 \else \expandafter \@secondoftwo
 \fi
}%
\providecommand \@ifx [1]{%
 \ifx #1\expandafter \@firstoftwo
 \else \expandafter \@secondoftwo
 \fi
}%
\providecommand \natexlab [1]{#1}%
\providecommand \enquote  [1]{``#1''}%
\providecommand \bibnamefont  [1]{#1}%
\providecommand \bibfnamefont [1]{#1}%
\providecommand \citenamefont [1]{#1}%
\providecommand \href@noop [0]{\@secondoftwo}%
\providecommand \href [0]{\begingroup \@sanitize@url \@href}%
\providecommand \@href[1]{\@@startlink{#1}\@@href}%
\providecommand \@@href[1]{\endgroup#1\@@endlink}%
\providecommand \@sanitize@url [0]{\catcode `\\12\catcode `\$12\catcode
  `\&12\catcode `\#12\catcode `\^12\catcode `\_12\catcode `\%12\relax}%
\providecommand \@@startlink[1]{}%
\providecommand \@@endlink[0]{}%
\providecommand \url  [0]{\begingroup\@sanitize@url \@url }%
\providecommand \@url [1]{\endgroup\@href {#1}{\urlprefix }}%
\providecommand \urlprefix  [0]{URL }%
\providecommand \Eprint [0]{\href }%
\providecommand \doibase [0]{https://doi.org/}%
\providecommand \selectlanguage [0]{\@gobble}%
\providecommand \bibinfo  [0]{\@secondoftwo}%
\providecommand \bibfield  [0]{\@secondoftwo}%
\providecommand \translation [1]{[#1]}%
\providecommand \BibitemOpen [0]{}%
\providecommand \bibitemStop [0]{}%
\providecommand \bibitemNoStop [0]{.\EOS\space}%
\providecommand \EOS [0]{\spacefactor3000\relax}%
\providecommand \BibitemShut  [1]{\csname bibitem#1\endcsname}%
\let\auto@bib@innerbib\@empty
\bibitem [{\citenamefont {Wen}(2019)}]{spinwen2019}%
  \BibitemOpen
  \bibfield  {author} {\bibinfo {author} {\bibfnamefont {X.-G.}\ \bibnamefont
  {Wen}},\ }\bibfield  {title} {\bibinfo {title} {Choreographed entanglement
  dances: Topological states of quantum matter},\ }\href
  {https://doi.org/10.1126/science.aal3099} {\bibfield  {journal} {\bibinfo
  {journal} {Science}\ }\textbf {\bibinfo {volume} {363}},\ \bibinfo {pages}
  {eaal3099} (\bibinfo {year} {2019})}\BibitemShut {NoStop}%
\bibitem [{\citenamefont {Gurevich}(2014)}]{gurevich2014}%
  \BibitemOpen
  \bibfield  {author} {\bibinfo {author} {\bibfnamefont {A.}~\bibnamefont
  {Gurevich}},\ }\bibfield  {title} {\bibinfo {title} {Challenges and
  opportunities for applications of unconventional superconductors},\ }\href
  {https://doi.org/10.1146/annurev-conmatphys-031113-133822} {\bibfield
  {journal} {\bibinfo  {journal} {Ann. Rev. Cond. Matter Phys.}\ }\textbf
  {\bibinfo {volume} {5}},\ \bibinfo {pages} {35} (\bibinfo {year}
  {2014})}\BibitemShut {NoStop}%
\bibitem [{\citenamefont {Cho}(2009)}]{cho2009}%
  \BibitemOpen
  \bibfield  {author} {\bibinfo {author} {\bibfnamefont {A.}~\bibnamefont
  {Cho}},\ }\bibfield  {title} {\bibinfo {title} {Helium-3 shortage could put
  freeze on low-temperature research},\ }\href
  {https://doi.org/10.1126/science.326_778} {\bibfield  {journal} {\bibinfo
  {journal} {Science}\ }\textbf {\bibinfo {volume} {326}},\ \bibinfo {pages}
  {778} (\bibinfo {year} {2009})}\BibitemShut {NoStop}%
\bibitem [{\citenamefont {Nuttall}\ \emph {et~al.}(2012)\citenamefont
  {Nuttall}, \citenamefont {Clarke},\ and\ \citenamefont
  {Glowacki}}]{nuttall2012}%
  \BibitemOpen
  \bibfield  {author} {\bibinfo {author} {\bibfnamefont {W.~J.}\ \bibnamefont
  {Nuttall}}, \bibinfo {author} {\bibfnamefont {R.~H.}\ \bibnamefont
  {Clarke}},\ and\ \bibinfo {author} {\bibfnamefont {B.~A.}\ \bibnamefont
  {Glowacki}},\ }\bibfield  {title} {\bibinfo {title} {Stop squandering
  helium},\ }\href {https://doi.org/10.1038/485573a} {\bibfield  {journal}
  {\bibinfo  {journal} {Nature}\ }\textbf {\bibinfo {volume} {485}},\ \bibinfo
  {pages} {573} (\bibinfo {year} {2012})}\BibitemShut {NoStop}%
\bibitem [{\citenamefont {Debye}(1926)}]{debye1926}%
  \BibitemOpen
  \bibfield  {author} {\bibinfo {author} {\bibfnamefont {P.}~\bibnamefont
  {Debye}},\ }\bibfield  {title} {\bibinfo {title} {{Einige Bemerkungen zur
  Magnetisierung bei tiefer Temperatur}},\ }\href
  {https://doi.org/10.1002/andp.19263862517} {\bibfield  {journal} {\bibinfo
  {journal} {Ann. Phys.}\ }\textbf {\bibinfo {volume} {386}},\ \bibinfo {pages}
  {1154} (\bibinfo {year} {1926})}\BibitemShut {NoStop}%
\bibitem [{\citenamefont {Giauque}(1927)}]{giauque1927}%
  \BibitemOpen
  \bibfield  {author} {\bibinfo {author} {\bibfnamefont {W.~F.}\ \bibnamefont
  {Giauque}},\ }\bibfield  {title} {\bibinfo {title} {A thermodynamic treatment
  of certain magnetic effects. a proposed method of producing temperatures
  considerably below $1^{\circ}$ absolute},\ }\href
  {https://doi.org/10.1021/ja01407a003} {\bibfield  {journal} {\bibinfo
  {journal} {J. Amer. Chem. Soc.}\ }\textbf {\bibinfo {volume} {49}},\ \bibinfo
  {pages} {1864} (\bibinfo {year} {1927})}\BibitemShut {NoStop}%
\bibitem [{\citenamefont {Wikus}\ \emph {et~al.}(2014)\citenamefont {Wikus},
  \citenamefont {Canavan}, \citenamefont {{Trowbridge Heine}}, \citenamefont
  {Matsumoto},\ and\ \citenamefont {Numazawa}}]{wikus2014}%
  \BibitemOpen
  \bibfield  {author} {\bibinfo {author} {\bibfnamefont {P.}~\bibnamefont
  {Wikus}}, \bibinfo {author} {\bibfnamefont {E.}~\bibnamefont {Canavan}},
  \bibinfo {author} {\bibfnamefont {S.}~\bibnamefont {{Trowbridge Heine}}},
  \bibinfo {author} {\bibfnamefont {K.}~\bibnamefont {Matsumoto}},\ and\
  \bibinfo {author} {\bibfnamefont {T.}~\bibnamefont {Numazawa}},\ }\bibfield
  {title} {\bibinfo {title} {Magnetocaloric materials and the optimization of
  cooling power density},\ }\href
  {https://doi.org/10.1016/j.cryogenics.2014.04.005} {\bibfield  {journal}
  {\bibinfo  {journal} {Cryogenics}\ }\textbf {\bibinfo {volume} {62}},\
  \bibinfo {pages} {150} (\bibinfo {year} {2014})}\BibitemShut {NoStop}%
\bibitem [{\citenamefont {Tokiwa}\ \emph {et~al.}(2021)\citenamefont {Tokiwa},
  \citenamefont {Bachus}, \citenamefont {Kavita}, \citenamefont {Jesche},
  \citenamefont {Tsirlin},\ and\ \citenamefont {Gegenwart}}]{Tokiwa2021}%
  \BibitemOpen
  \bibfield  {author} {\bibinfo {author} {\bibfnamefont {Y.}~\bibnamefont
  {Tokiwa}}, \bibinfo {author} {\bibfnamefont {S.}~\bibnamefont {Bachus}},
  \bibinfo {author} {\bibfnamefont {K.}~\bibnamefont {Kavita}}, \bibinfo
  {author} {\bibfnamefont {A.}~\bibnamefont {Jesche}}, \bibinfo {author}
  {\bibfnamefont {A.~A.}\ \bibnamefont {Tsirlin}},\ and\ \bibinfo {author}
  {\bibfnamefont {P.}~\bibnamefont {Gegenwart}},\ }\bibfield  {title} {\bibinfo
  {title} {{Frustrated magnet for adiabatic demagnetization cooling to
  milli-Kelvin temperatures}},\ }\bibfield  {journal} {\bibinfo  {journal}
  {Commun. Mater.}\ }\textbf {\bibinfo {volume} {2}},\ \href
  {https://doi.org/10.1038/s43246-021-00142-1} {10.1038/s43246-021-00142-1}
  (\bibinfo {year} {2021})\BibitemShut {NoStop}%
\bibitem [{\citenamefont {Guo}\ \emph {et~al.}(2019{\natexlab{a}})\citenamefont
  {Guo}, \citenamefont {Kong}, \citenamefont {{Alex Cevallos}}, \citenamefont
  {Stolze},\ and\ \citenamefont {Cava}}]{guo2019}%
  \BibitemOpen
  \bibfield  {author} {\bibinfo {author} {\bibfnamefont {S.}~\bibnamefont
  {Guo}}, \bibinfo {author} {\bibfnamefont {T.}~\bibnamefont {Kong}}, \bibinfo
  {author} {\bibfnamefont {F.}~\bibnamefont {{Alex Cevallos}}}, \bibinfo
  {author} {\bibfnamefont {K.}~\bibnamefont {Stolze}},\ and\ \bibinfo {author}
  {\bibfnamefont {R.}~\bibnamefont {Cava}},\ }\bibfield  {title} {\bibinfo
  {title} {Crystal growth, crystal structure and anisotropic magnetic
  properties of {KBaR(BO$_3)_2$ (R = Y, Gd, Tb, Dy, Ho, Tm, Yb, and Lu)}
  triangular lattice materials},\ }\href
  {https://doi.org/10.1016/j.jmmm.2018.10.037} {\bibfield  {journal} {\bibinfo
  {journal} {J. Magn. Magn. Mater.}\ }\textbf {\bibinfo {volume} {472}},\
  \bibinfo {pages} {104} (\bibinfo {year} {2019}{\natexlab{a}})}\BibitemShut
  {NoStop}%
\bibitem [{\citenamefont {Li}\ \emph {et~al.}(2017)\citenamefont {Li},
  \citenamefont {Adroja}, \citenamefont {Bewley}, \citenamefont {Voneshen},
  \citenamefont {Tsirlin}, \citenamefont {Gegenwart},\ and\ \citenamefont
  {Zhang}}]{li2017}%
  \BibitemOpen
  \bibfield  {author} {\bibinfo {author} {\bibfnamefont {Y.}~\bibnamefont
  {Li}}, \bibinfo {author} {\bibfnamefont {D.}~\bibnamefont {Adroja}}, \bibinfo
  {author} {\bibfnamefont {R.~I.}\ \bibnamefont {Bewley}}, \bibinfo {author}
  {\bibfnamefont {D.}~\bibnamefont {Voneshen}}, \bibinfo {author}
  {\bibfnamefont {A.~A.}\ \bibnamefont {Tsirlin}}, \bibinfo {author}
  {\bibfnamefont {P.}~\bibnamefont {Gegenwart}},\ and\ \bibinfo {author}
  {\bibfnamefont {Q.}~\bibnamefont {Zhang}},\ }\bibfield  {title} {\bibinfo
  {title} {Crystalline electric-field randomness in the triangular lattice
  spin-liquid {YbMgGaO$_4$}},\ }\href
  {https://doi.org/10.1103/PhysRevLett.118.107202} {\bibfield  {journal}
  {\bibinfo  {journal} {Phys. Rev. Lett.}\ }\textbf {\bibinfo {volume} {118}},\
  \bibinfo {pages} {107202} (\bibinfo {year} {2017})}\BibitemShut {NoStop}%
\bibitem [{\citenamefont {Rau}\ and\ \citenamefont {Gingras}(2018)}]{rau2018}%
  \BibitemOpen
  \bibfield  {author} {\bibinfo {author} {\bibfnamefont {J.~G.}\ \bibnamefont
  {Rau}}\ and\ \bibinfo {author} {\bibfnamefont {M.~J.~P.}\ \bibnamefont
  {Gingras}},\ }\bibfield  {title} {\bibinfo {title} {Frustration and
  anisotropic exchange in ytterbium magnets with edge-shared octahedra},\
  }\href {https://doi.org/10.1103/PhysRevB.98.054408} {\bibfield  {journal}
  {\bibinfo  {journal} {Phys. Rev. B}\ }\textbf {\bibinfo {volume} {98}},\
  \bibinfo {pages} {054408} (\bibinfo {year} {2018})}\BibitemShut {NoStop}%
\bibitem [{\citenamefont {Li}\ \emph {et~al.}(2020)\citenamefont {Li},
  \citenamefont {Gegenwart},\ and\ \citenamefont {Tsirlin}}]{li2020}%
  \BibitemOpen
  \bibfield  {author} {\bibinfo {author} {\bibfnamefont {Y.}~\bibnamefont
  {Li}}, \bibinfo {author} {\bibfnamefont {P.}~\bibnamefont {Gegenwart}},\ and\
  \bibinfo {author} {\bibfnamefont {A.~A.}\ \bibnamefont {Tsirlin}},\
  }\bibfield  {title} {\bibinfo {title} {Spin liquids in geometrically perfect
  triangular antiferromagnets},\ }\href
  {https://doi.org/10.1088/1361-648X/ab724e} {\bibfield  {journal} {\bibinfo
  {journal} {J. Phys.: Condens. Matter}\ }\textbf {\bibinfo {volume} {32}},\
  \bibinfo {pages} {224004} (\bibinfo {year} {2020})}\BibitemShut {NoStop}%
\bibitem [{\citenamefont {Sanders}\ \emph {et~al.}(2017)\citenamefont
  {Sanders}, \citenamefont {Cevallos},\ and\ \citenamefont
  {Cava}}]{sanders2017}%
  \BibitemOpen
  \bibfield  {author} {\bibinfo {author} {\bibfnamefont {M.~B.}\ \bibnamefont
  {Sanders}}, \bibinfo {author} {\bibfnamefont {F.~A.}\ \bibnamefont
  {Cevallos}},\ and\ \bibinfo {author} {\bibfnamefont {R.~J.}\ \bibnamefont
  {Cava}},\ }\bibfield  {title} {\bibinfo {title} {Magnetism in the
  {KBaRE(BO$_3)_2$ (RE = Sm, Eu, Gd, Tb, Dy, Ho, Er, Tm, Yb, Lu)} series:
  materials with a triangular rare earth lattice},\ }\href
  {https://doi.org/10.1088/2053-1591/aa60a2} {\bibfield  {journal} {\bibinfo
  {journal} {Mater. Res. Express}\ }\textbf {\bibinfo {volume} {4}},\ \bibinfo
  {pages} {036102} (\bibinfo {year} {2017})}\BibitemShut {NoStop}%
\bibitem [{Uti()}]{Utility2020}%
  \BibitemOpen
  \href@noop {} {}\bibinfo {note} {P. Gegenwart, A.A. Tsirlin, S. Bachus, Y.
  Tokiwa, DE: 20 2020 002 079.6 (12.05.2020), IPC: C09K 5/06}\BibitemShut
  {NoStop}%
\bibitem [{\citenamefont {Pet{\u r}{\'\i}{\u c}ek}\ \emph
  {et~al.}(2014)\citenamefont {Pet{\u r}{\'\i}{\u c}ek}, \citenamefont {Du{\u
  s}ek},\ and\ \citenamefont {Palatinus}}]{Petricek2014}%
  \BibitemOpen
  \bibfield  {author} {\bibinfo {author} {\bibfnamefont {V.}~\bibnamefont
  {Pet{\u r}{\'\i}{\u c}ek}}, \bibinfo {author} {\bibfnamefont
  {M.}~\bibnamefont {Du{\u s}ek}},\ and\ \bibinfo {author} {\bibfnamefont
  {L.}~\bibnamefont {Palatinus}},\ }\bibfield  {title} {\bibinfo {title}
  {Crystallographic computing system {JANA2006}: General features},\ }\href
  {https://doi.org/10.1515/zkri-2014-1737} {\bibfield  {journal} {\bibinfo
  {journal} {Z. Krist.}\ }\textbf {\bibinfo {volume} {229}},\ \bibinfo {pages}
  {345} (\bibinfo {year} {2014})}\BibitemShut {NoStop}%
\bibitem [{\citenamefont {Guo}\ \emph {et~al.}(2019{\natexlab{b}})\citenamefont
  {Guo}, \citenamefont {Ghasemi}, \citenamefont {Broholm},\ and\ \citenamefont
  {Cava}}]{guo2019b}%
  \BibitemOpen
  \bibfield  {author} {\bibinfo {author} {\bibfnamefont {S.}~\bibnamefont
  {Guo}}, \bibinfo {author} {\bibfnamefont {A.}~\bibnamefont {Ghasemi}},
  \bibinfo {author} {\bibfnamefont {C.~L.}\ \bibnamefont {Broholm}},\ and\
  \bibinfo {author} {\bibfnamefont {R.~J.}\ \bibnamefont {Cava}},\ }\bibfield
  {title} {\bibinfo {title} {Magnetism on ideal triangular lattices in
  {NaBaYb(BO$_3)_2$}},\ }\href
  {https://doi.org/10.1103/PhysRevMaterials.3.094404} {\bibfield  {journal}
  {\bibinfo  {journal} {Phys. Rev. Materials}\ }\textbf {\bibinfo {volume}
  {3}},\ \bibinfo {pages} {094404} (\bibinfo {year}
  {2019}{\natexlab{b}})}\BibitemShut {NoStop}%
\bibitem [{\citenamefont {Guo}\ \emph {et~al.}(2019{\natexlab{c}})\citenamefont
  {Guo}, \citenamefont {Kong}, \citenamefont {Xie}, \citenamefont {Nguyen},
  \citenamefont {Stolze}, \citenamefont {{Alex Cevallos}},\ and\ \citenamefont
  {Cava}}]{guo2019c}%
  \BibitemOpen
  \bibfield  {author} {\bibinfo {author} {\bibfnamefont {S.}~\bibnamefont
  {Guo}}, \bibinfo {author} {\bibfnamefont {T.}~\bibnamefont {Kong}}, \bibinfo
  {author} {\bibfnamefont {W.}~\bibnamefont {Xie}}, \bibinfo {author}
  {\bibfnamefont {L.}~\bibnamefont {Nguyen}}, \bibinfo {author} {\bibfnamefont
  {K.}~\bibnamefont {Stolze}}, \bibinfo {author} {\bibfnamefont
  {F.}~\bibnamefont {{Alex Cevallos}}},\ and\ \bibinfo {author} {\bibfnamefont
  {R.~J.}\ \bibnamefont {Cava}},\ }\bibfield  {title} {\bibinfo {title}
  {Triangular rare-earth lattice materials {RbBaR(BO$_3)_2$ (R = Y, Gd-Yb)} and
  comparison to the {KBaR(BO3)$_2$} analogs},\ }\href
  {https://doi.org/10.1021/acs.inorgchem.8b03372} {\bibfield  {journal}
  {\bibinfo  {journal} {Inorg. Chem.}\ }\textbf {\bibinfo {volume} {58}},\
  \bibinfo {pages} {3308} (\bibinfo {year} {2019}{\natexlab{c}})}\BibitemShut
  {NoStop}%
\bibitem [{\citenamefont {Langford}\ and\ \citenamefont
  {Wilson}(1978)}]{Langford1978}%
  \BibitemOpen
  \bibfield  {author} {\bibinfo {author} {\bibfnamefont {J.~I.}\ \bibnamefont
  {Langford}}\ and\ \bibinfo {author} {\bibfnamefont {A.~J.~C.}\ \bibnamefont
  {Wilson}},\ }\bibfield  {title} {\bibinfo {title} {Scherrer after sixty
  years: A survey and some new results in the determination of crystallite
  size},\ }\href {https://doi.org/10.1107/S0021889878012844} {\bibfield
  {journal} {\bibinfo  {journal} {J. Appl. Cryst.}\ }\textbf {\bibinfo {volume}
  {11}},\ \bibinfo {pages} {102} (\bibinfo {year} {1978})}\BibitemShut
  {NoStop}%
\bibitem [{\citenamefont {Johnston}\ \emph {et~al.}(2000)\citenamefont
  {Johnston}, \citenamefont {Kremer}, \citenamefont {Troyer}, \citenamefont
  {Wang}, \citenamefont {Kl\"umper}, \citenamefont {Bud'ko}, \citenamefont
  {Panchula},\ and\ \citenamefont {Canfield}}]{Johnston2000}%
  \BibitemOpen
  \bibfield  {author} {\bibinfo {author} {\bibfnamefont {D.~C.}\ \bibnamefont
  {Johnston}}, \bibinfo {author} {\bibfnamefont {R.~K.}\ \bibnamefont
  {Kremer}}, \bibinfo {author} {\bibfnamefont {M.}~\bibnamefont {Troyer}},
  \bibinfo {author} {\bibfnamefont {X.}~\bibnamefont {Wang}}, \bibinfo {author}
  {\bibfnamefont {A.}~\bibnamefont {Kl\"umper}}, \bibinfo {author}
  {\bibfnamefont {S.~L.}\ \bibnamefont {Bud'ko}}, \bibinfo {author}
  {\bibfnamefont {A.~F.}\ \bibnamefont {Panchula}},\ and\ \bibinfo {author}
  {\bibfnamefont {P.~C.}\ \bibnamefont {Canfield}},\ }\bibfield  {title}
  {\bibinfo {title} {{Thermodynamics of spin $S=1/2$ antiferromagnetic uniform
  and alternating-exchange Heisenberg chains}},\ }\href
  {https://doi.org/10.1103/PhysRevB.61.9558} {\bibfield  {journal} {\bibinfo
  {journal} {Phys. Rev. B}\ }\textbf {\bibinfo {volume} {61}},\ \bibinfo
  {pages} {9558} (\bibinfo {year} {2000})}\BibitemShut {NoStop}%
\bibitem [{\citenamefont {Zhu}\ \emph {et~al.}(2017)\citenamefont {Zhu},
  \citenamefont {Maksimov}, \citenamefont {White},\ and\ \citenamefont
  {Chernyshev}}]{Zhu2017}%
  \BibitemOpen
  \bibfield  {author} {\bibinfo {author} {\bibfnamefont {Z.}~\bibnamefont
  {Zhu}}, \bibinfo {author} {\bibfnamefont {P.~A.}\ \bibnamefont {Maksimov}},
  \bibinfo {author} {\bibfnamefont {S.~R.}\ \bibnamefont {White}},\ and\
  \bibinfo {author} {\bibfnamefont {A.~L.}\ \bibnamefont {Chernyshev}},\
  }\bibfield  {title} {\bibinfo {title} {{Disorder-Induced Mimicry of a Spin
  Liquid in ${\mathrm{YbMgGaO}}_{4}$}},\ }\href
  {https://doi.org/10.1103/PhysRevLett.119.157201} {\bibfield  {journal}
  {\bibinfo  {journal} {Phys. Rev. Lett.}\ }\textbf {\bibinfo {volume} {119}},\
  \bibinfo {pages} {157201} (\bibinfo {year} {2017})}\BibitemShut {NoStop}%
\bibitem [{\citenamefont {Svistov}\ \emph {et~al.}(2006)\citenamefont
  {Svistov}, \citenamefont {Smirnov}, \citenamefont {Prozorova}, \citenamefont
  {Petrenko}, \citenamefont {Micheler}, \citenamefont {B\"uttgen},
  \citenamefont {Shapiro},\ and\ \citenamefont {Demianets}}]{svistov2006}%
  \BibitemOpen
  \bibfield  {author} {\bibinfo {author} {\bibfnamefont {L.~E.}\ \bibnamefont
  {Svistov}}, \bibinfo {author} {\bibfnamefont {A.~I.}\ \bibnamefont
  {Smirnov}}, \bibinfo {author} {\bibfnamefont {L.~A.}\ \bibnamefont
  {Prozorova}}, \bibinfo {author} {\bibfnamefont {O.~A.}\ \bibnamefont
  {Petrenko}}, \bibinfo {author} {\bibfnamefont {A.}~\bibnamefont {Micheler}},
  \bibinfo {author} {\bibfnamefont {N.}~\bibnamefont {B\"uttgen}}, \bibinfo
  {author} {\bibfnamefont {A.~Y.}\ \bibnamefont {Shapiro}},\ and\ \bibinfo
  {author} {\bibfnamefont {L.~N.}\ \bibnamefont {Demianets}},\ }\bibfield
  {title} {\bibinfo {title} {{Magnetic phase diagram, critical behavior, and
  two-dimensional to three-dimensional crossover in the triangular lattice
  antiferromagnet {RbFe(MoO$_4$)$_2$}}},\ }\href
  {https://doi.org/10.1103/PhysRevB.74.024412} {\bibfield  {journal} {\bibinfo
  {journal} {Phys. Rev. B}\ }\textbf {\bibinfo {volume} {74}},\ \bibinfo
  {pages} {024412} (\bibinfo {year} {2006})}\BibitemShut {NoStop}%
\bibitem [{\citenamefont {Smirnov}\ \emph {et~al.}(2009)\citenamefont
  {Smirnov}, \citenamefont {Svistov}, \citenamefont {Prozorova}, \citenamefont
  {Zheludev}, \citenamefont {Lumsden}, \citenamefont {Ressouche}, \citenamefont
  {Petrenko}, \citenamefont {Nishikawa}, \citenamefont {Kimura}, \citenamefont
  {Hagiwara}, \citenamefont {Kindo}, \citenamefont {Shapiro},\ and\
  \citenamefont {Demianets}}]{smirnov2009}%
  \BibitemOpen
  \bibfield  {author} {\bibinfo {author} {\bibfnamefont {A.~I.}\ \bibnamefont
  {Smirnov}}, \bibinfo {author} {\bibfnamefont {L.~E.}\ \bibnamefont
  {Svistov}}, \bibinfo {author} {\bibfnamefont {L.~A.}\ \bibnamefont
  {Prozorova}}, \bibinfo {author} {\bibfnamefont {A.}~\bibnamefont {Zheludev}},
  \bibinfo {author} {\bibfnamefont {M.~D.}\ \bibnamefont {Lumsden}}, \bibinfo
  {author} {\bibfnamefont {E.}~\bibnamefont {Ressouche}}, \bibinfo {author}
  {\bibfnamefont {O.~A.}\ \bibnamefont {Petrenko}}, \bibinfo {author}
  {\bibfnamefont {K.}~\bibnamefont {Nishikawa}}, \bibinfo {author}
  {\bibfnamefont {S.}~\bibnamefont {Kimura}}, \bibinfo {author} {\bibfnamefont
  {M.}~\bibnamefont {Hagiwara}}, \bibinfo {author} {\bibfnamefont
  {K.}~\bibnamefont {Kindo}}, \bibinfo {author} {\bibfnamefont {A.~Y.}\
  \bibnamefont {Shapiro}},\ and\ \bibinfo {author} {\bibfnamefont {L.~N.}\
  \bibnamefont {Demianets}},\ }\bibfield  {title} {\bibinfo {title} {Chiral and
  collinear ordering in a distorted triangular antiferromagnet},\ }\href
  {https://doi.org/10.1103/PhysRevLett.102.037202} {\bibfield  {journal}
  {\bibinfo  {journal} {Phys. Rev. Lett.}\ }\textbf {\bibinfo {volume} {102}},\
  \bibinfo {pages} {037202} (\bibinfo {year} {2009})}\BibitemShut {NoStop}%
\bibitem [{\citenamefont {Lebernegg}\ \emph {et~al.}(2016)\citenamefont
  {Lebernegg}, \citenamefont {Tsirlin}, \citenamefont {Janson}, \citenamefont
  {Redhammer},\ and\ \citenamefont {Rosner}}]{lebernegg2016}%
  \BibitemOpen
  \bibfield  {author} {\bibinfo {author} {\bibfnamefont {S.}~\bibnamefont
  {Lebernegg}}, \bibinfo {author} {\bibfnamefont {A.~A.}\ \bibnamefont
  {Tsirlin}}, \bibinfo {author} {\bibfnamefont {O.}~\bibnamefont {Janson}},
  \bibinfo {author} {\bibfnamefont {G.~J.}\ \bibnamefont {Redhammer}},\ and\
  \bibinfo {author} {\bibfnamefont {H.}~\bibnamefont {Rosner}},\ }\bibfield
  {title} {\bibinfo {title} {Interplay of magnetic sublattices in langite
  {Cu$_4$(OH)$_6$SO$_4\cdot 2$H$_2$O}},\ }\href
  {https://doi.org/10.1088/1367-2630/18/3/033020} {\bibfield  {journal}
  {\bibinfo  {journal} {New J. Phys.}\ }\textbf {\bibinfo {volume} {18}},\
  \bibinfo {pages} {033020} (\bibinfo {year} {2016})}\BibitemShut {NoStop}%
\bibitem [{\citenamefont {Somesh}\ \emph {et~al.}(2021)\citenamefont {Somesh},
  \citenamefont {Furukawa}, \citenamefont {Simutis}, \citenamefont {Bert},
  \citenamefont {Prinz-Zwick}, \citenamefont {B\"uttgen}, \citenamefont
  {Zorko}, \citenamefont {Tsirlin}, \citenamefont {Mendels},\ and\
  \citenamefont {Nath}}]{somesh2021}%
  \BibitemOpen
  \bibfield  {author} {\bibinfo {author} {\bibfnamefont {K.}~\bibnamefont
  {Somesh}}, \bibinfo {author} {\bibfnamefont {Y.}~\bibnamefont {Furukawa}},
  \bibinfo {author} {\bibfnamefont {G.}~\bibnamefont {Simutis}}, \bibinfo
  {author} {\bibfnamefont {F.}~\bibnamefont {Bert}}, \bibinfo {author}
  {\bibfnamefont {M.}~\bibnamefont {Prinz-Zwick}}, \bibinfo {author}
  {\bibfnamefont {N.}~\bibnamefont {B\"uttgen}}, \bibinfo {author}
  {\bibfnamefont {A.}~\bibnamefont {Zorko}}, \bibinfo {author} {\bibfnamefont
  {A.~A.}\ \bibnamefont {Tsirlin}}, \bibinfo {author} {\bibfnamefont
  {P.}~\bibnamefont {Mendels}},\ and\ \bibinfo {author} {\bibfnamefont
  {R.}~\bibnamefont {Nath}},\ }\bibfield  {title} {\bibinfo {title} {Universal
  fluctuating regime in triangular chromate antiferromagnets},\ }\href
  {https://doi.org/10.1103/PhysRevB.104.104422} {\bibfield  {journal} {\bibinfo
   {journal} {Phys. Rev. B}\ }\textbf {\bibinfo {volume} {104}},\ \bibinfo
  {pages} {104422} (\bibinfo {year} {2021})}\BibitemShut {NoStop}%
\bibitem [{\citenamefont {Bag}\ \emph {et~al.}(2021)\citenamefont {Bag},
  \citenamefont {Ennis}, \citenamefont {Liu}, \citenamefont {Dissanayake},
  \citenamefont {Shi}, \citenamefont {Liu}, \citenamefont {Balents},\ and\
  \citenamefont {Haravifard}}]{bag2021}%
  \BibitemOpen
  \bibfield  {author} {\bibinfo {author} {\bibfnamefont {R.}~\bibnamefont
  {Bag}}, \bibinfo {author} {\bibfnamefont {M.}~\bibnamefont {Ennis}}, \bibinfo
  {author} {\bibfnamefont {C.}~\bibnamefont {Liu}}, \bibinfo {author}
  {\bibfnamefont {S.~E.}\ \bibnamefont {Dissanayake}}, \bibinfo {author}
  {\bibfnamefont {Z.}~\bibnamefont {Shi}}, \bibinfo {author} {\bibfnamefont
  {J.}~\bibnamefont {Liu}}, \bibinfo {author} {\bibfnamefont {L.}~\bibnamefont
  {Balents}},\ and\ \bibinfo {author} {\bibfnamefont {S.}~\bibnamefont
  {Haravifard}},\ }\bibfield  {title} {\bibinfo {title} {Realization of quantum
  dipoles in triangular lattice crystal {Ba$_3$Yb(BO$_3)_3$}},\ }\href
  {https://doi.org/10.1103/PhysRevB.104.L220403} {\bibfield  {journal}
  {\bibinfo  {journal} {Phys. Rev. B}\ }\textbf {\bibinfo {volume} {104}},\
  \bibinfo {pages} {L220403} (\bibinfo {year} {2021})}\BibitemShut {NoStop}%
\bibitem [{\citenamefont {Jiang}\ \emph {et~al.}(2022)\citenamefont {Jiang},
  \citenamefont {Yang}, \citenamefont {Gao}, \citenamefont {Wan}, \citenamefont
  {Zhu}, \citenamefont {Shiroka}, \citenamefont {Chen}, \citenamefont {Wu},
  \citenamefont {Li}, \citenamefont {Jiao}, \citenamefont {Chen}, \citenamefont
  {Bao}, \citenamefont {Tian},\ and\ \citenamefont {Shu}}]{jiang2022}%
  \BibitemOpen
  \bibfield  {author} {\bibinfo {author} {\bibfnamefont {C.~Y.}\ \bibnamefont
  {Jiang}}, \bibinfo {author} {\bibfnamefont {Y.~X.}\ \bibnamefont {Yang}},
  \bibinfo {author} {\bibfnamefont {Y.~X.}\ \bibnamefont {Gao}}, \bibinfo
  {author} {\bibfnamefont {Z.~T.}\ \bibnamefont {Wan}}, \bibinfo {author}
  {\bibfnamefont {Z.~H.}\ \bibnamefont {Zhu}}, \bibinfo {author} {\bibfnamefont
  {T.}~\bibnamefont {Shiroka}}, \bibinfo {author} {\bibfnamefont {C.~S.}\
  \bibnamefont {Chen}}, \bibinfo {author} {\bibfnamefont {Q.}~\bibnamefont
  {Wu}}, \bibinfo {author} {\bibfnamefont {X.}~\bibnamefont {Li}}, \bibinfo
  {author} {\bibfnamefont {J.~C.}\ \bibnamefont {Jiao}}, \bibinfo {author}
  {\bibfnamefont {K.~W.}\ \bibnamefont {Chen}}, \bibinfo {author}
  {\bibfnamefont {Y.}~\bibnamefont {Bao}}, \bibinfo {author} {\bibfnamefont
  {Z.~M.}\ \bibnamefont {Tian}},\ and\ \bibinfo {author} {\bibfnamefont
  {L.}~\bibnamefont {Shu}},\ }\bibfield  {title} {\bibinfo {title} {Spin
  excitations in the quantum dipolar magnet {Yb(BaBO$_3)_3$}},\ }\href
  {https://doi.org/10.1103/PhysRevB.106.014409} {\bibfield  {journal} {\bibinfo
   {journal} {Phys. Rev. B}\ }\textbf {\bibinfo {volume} {106}},\ \bibinfo
  {pages} {014409} (\bibinfo {year} {2022})}\BibitemShut {NoStop}%
\bibitem [{\citenamefont {Huse}\ and\ \citenamefont {Elser}(1988)}]{huse1988}%
  \BibitemOpen
  \bibfield  {author} {\bibinfo {author} {\bibfnamefont {D.~A.}\ \bibnamefont
  {Huse}}\ and\ \bibinfo {author} {\bibfnamefont {V.}~\bibnamefont {Elser}},\
  }\bibfield  {title} {\bibinfo {title} {{Simple Variational Wave Functions for
  Two-Dimensional Heisenberg Spin-$\frac12$ Antiferromagnets}},\ }\href
  {https://doi.org/10.1103/PhysRevLett.60.2531} {\bibfield  {journal} {\bibinfo
   {journal} {Phys. Rev. Lett.}\ }\textbf {\bibinfo {volume} {60}},\ \bibinfo
  {pages} {2531} (\bibinfo {year} {1988})}\BibitemShut {NoStop}%
\bibitem [{\citenamefont {Kele\c{s}}\ and\ \citenamefont
  {Zhao}(2018)}]{keles2018}%
  \BibitemOpen
  \bibfield  {author} {\bibinfo {author} {\bibfnamefont {A.}~\bibnamefont
  {Kele\c{s}}}\ and\ \bibinfo {author} {\bibfnamefont {E.}~\bibnamefont
  {Zhao}},\ }\bibfield  {title} {\bibinfo {title} {Absence of long-range order
  in a triangular spin system with dipolar interactions},\ }\href
  {https://doi.org/10.1103/PhysRevLett.120.187202} {\bibfield  {journal}
  {\bibinfo  {journal} {Phys. Rev. Lett.}\ }\textbf {\bibinfo {volume} {120}},\
  \bibinfo {pages} {187202} (\bibinfo {year} {2018})}\BibitemShut {NoStop}%
\bibitem [{\citenamefont {Vilches}\ and\ \citenamefont
  {Wheatley}(1966)}]{vilches1966}%
  \BibitemOpen
  \bibfield  {author} {\bibinfo {author} {\bibfnamefont {O.~E.}\ \bibnamefont
  {Vilches}}\ and\ \bibinfo {author} {\bibfnamefont {J.~C.}\ \bibnamefont
  {Wheatley}},\ }\bibfield  {title} {\bibinfo {title} {Measurements of the
  specific heats of three magnetic salts at low temperatures},\ }\href
  {https://doi.org/10.1103/PhysRev.148.509} {\bibfield  {journal} {\bibinfo
  {journal} {Phys. Rev.}\ }\textbf {\bibinfo {volume} {148}},\ \bibinfo {pages}
  {509} (\bibinfo {year} {1966})}\BibitemShut {NoStop}%
\bibitem [{\citenamefont {Jang}\ \emph {et~al.}(2015)\citenamefont {Jang},
  \citenamefont {Gruner}, \citenamefont {Steppke}, \citenamefont {Mitsumoto},
  \citenamefont {Geibel},\ and\ \citenamefont {Brando}}]{jang2015}%
  \BibitemOpen
  \bibfield  {author} {\bibinfo {author} {\bibfnamefont {D.}~\bibnamefont
  {Jang}}, \bibinfo {author} {\bibfnamefont {T.}~\bibnamefont {Gruner}},
  \bibinfo {author} {\bibfnamefont {A.}~\bibnamefont {Steppke}}, \bibinfo
  {author} {\bibfnamefont {K.}~\bibnamefont {Mitsumoto}}, \bibinfo {author}
  {\bibfnamefont {C.}~\bibnamefont {Geibel}},\ and\ \bibinfo {author}
  {\bibfnamefont {M.}~\bibnamefont {Brando}},\ }\bibfield  {title} {\bibinfo
  {title} {Large magnetocaloric effect and adiabatic demagnetization
  refrigeration with {YbPt$_2$Sn}},\ }\href
  {https://doi.org/10.1038/ncomms9680} {\bibfield  {journal} {\bibinfo
  {journal} {Nat. Commun.}\ }\textbf {\bibinfo {volume} {6}},\ \bibinfo {pages}
  {8680} (\bibinfo {year} {2015})}\BibitemShut {NoStop}%
\bibitem [{\citenamefont {Shimura}\ \emph {et~al.}(2022)\citenamefont
  {Shimura}, \citenamefont {Watanabe}, \citenamefont {Taniguchi}, \citenamefont
  {Osato}, \citenamefont {Yamamoto}, \citenamefont {Kusanose}, \citenamefont
  {Umeo}, \citenamefont {Fujita}, \citenamefont {Onimaru},\ and\ \citenamefont
  {Takabatake}}]{Shimura2022}%
  \BibitemOpen
  \bibfield  {author} {\bibinfo {author} {\bibfnamefont {Y.}~\bibnamefont
  {Shimura}}, \bibinfo {author} {\bibfnamefont {K.}~\bibnamefont {Watanabe}},
  \bibinfo {author} {\bibfnamefont {T.}~\bibnamefont {Taniguchi}}, \bibinfo
  {author} {\bibfnamefont {K.}~\bibnamefont {Osato}}, \bibinfo {author}
  {\bibfnamefont {R.}~\bibnamefont {Yamamoto}}, \bibinfo {author}
  {\bibfnamefont {Y.}~\bibnamefont {Kusanose}}, \bibinfo {author}
  {\bibfnamefont {K.}~\bibnamefont {Umeo}}, \bibinfo {author} {\bibfnamefont
  {M.}~\bibnamefont {Fujita}}, \bibinfo {author} {\bibfnamefont
  {T.}~\bibnamefont {Onimaru}},\ and\ \bibinfo {author} {\bibfnamefont
  {T.}~\bibnamefont {Takabatake}},\ }\bibfield  {title} {\bibinfo {title}
  {Magnetic refrigeration down to 0.2\,k by heavy fermion ybcu$_4$ni},\
  }\href@noop {} {\bibfield  {journal} {\bibinfo  {journal} {J. Appl. Phys.}\
  }\textbf {\bibinfo {volume} {131}},\ \bibinfo {pages} {013903} (\bibinfo
  {year} {2022})}\BibitemShut {NoStop}%
\bibitem [{\citenamefont {Kleinhans}\ \emph {et~al.}(2023)\citenamefont
  {Kleinhans}, \citenamefont {Eibensteiner}, \citenamefont {Leiner},
  \citenamefont {Resch}, \citenamefont {Worch}, \citenamefont {Wilde},
  \citenamefont {Spallek}, \citenamefont {Regnat},\ and\ \citenamefont
  {Pfleiderer}}]{Kleinhans2022}%
  \BibitemOpen
  \bibfield  {author} {\bibinfo {author} {\bibfnamefont {M.}~\bibnamefont
  {Kleinhans}}, \bibinfo {author} {\bibfnamefont {K.}~\bibnamefont
  {Eibensteiner}}, \bibinfo {author} {\bibfnamefont {J.}~\bibnamefont
  {Leiner}}, \bibinfo {author} {\bibfnamefont {C.}~\bibnamefont {Resch}},
  \bibinfo {author} {\bibfnamefont {L.}~\bibnamefont {Worch}}, \bibinfo
  {author} {\bibfnamefont {M.}~\bibnamefont {Wilde}}, \bibinfo {author}
  {\bibfnamefont {J.}~\bibnamefont {Spallek}}, \bibinfo {author} {\bibfnamefont
  {A.}~\bibnamefont {Regnat}},\ and\ \bibinfo {author} {\bibfnamefont
  {C.}~\bibnamefont {Pfleiderer}},\ }\bibfield  {title} {\bibinfo {title}
  {{Magnetocaloric Properties of ${R}_{3}{\mathrm{Ga}}_{5}{\mathrm{O}}_{12}$
  ($R=\text{Tb, Gd, Nd, Dy}$)}},\ }\href
  {https://doi.org/10.1103/PhysRevApplied.19.014038} {\bibfield  {journal}
  {\bibinfo  {journal} {Phys. Rev. Appl.}\ }\textbf {\bibinfo {volume} {19}},\
  \bibinfo {pages} {014038} (\bibinfo {year} {2023})}\BibitemShut {NoStop}%
\bibitem [{\citenamefont {Xiang}\ \emph {et~al.}(2023)\citenamefont {Xiang},
  \citenamefont {Su}, \citenamefont {Xi}, \citenamefont {Fu}, \citenamefont
  {Chen}, \citenamefont {Jin}, \citenamefont {Chen}, \citenamefont {Mo},
  \citenamefont {Qi}, \citenamefont {Shen}, \citenamefont {Zhang},
  \citenamefont {Jin}, \citenamefont {Li}, \citenamefont {Sun},\ and\
  \citenamefont {Su}}]{Xiang2023-arxiv}%
  \BibitemOpen
  \bibfield  {author} {\bibinfo {author} {\bibfnamefont {J.}~\bibnamefont
  {Xiang}}, \bibinfo {author} {\bibfnamefont {C.}~\bibnamefont {Su}}, \bibinfo
  {author} {\bibfnamefont {N.}~\bibnamefont {Xi}}, \bibinfo {author}
  {\bibfnamefont {Z.}~\bibnamefont {Fu}}, \bibinfo {author} {\bibfnamefont
  {Z.}~\bibnamefont {Chen}}, \bibinfo {author} {\bibfnamefont {H.}~\bibnamefont
  {Jin}}, \bibinfo {author} {\bibfnamefont {Z.}~\bibnamefont {Chen}}, \bibinfo
  {author} {\bibfnamefont {Z.-J.}\ \bibnamefont {Mo}}, \bibinfo {author}
  {\bibfnamefont {Y.}~\bibnamefont {Qi}}, \bibinfo {author} {\bibfnamefont
  {J.}~\bibnamefont {Shen}}, \bibinfo {author} {\bibfnamefont {L.}~\bibnamefont
  {Zhang}}, \bibinfo {author} {\bibfnamefont {W.}~\bibnamefont {Jin}}, \bibinfo
  {author} {\bibfnamefont {W.}~\bibnamefont {Li}}, \bibinfo {author}
  {\bibfnamefont {P.}~\bibnamefont {Sun}},\ and\ \bibinfo {author}
  {\bibfnamefont {G.}~\bibnamefont {Su}},\ }\href
  {https://doi.org/10.48550/ARXIV.2301.03571} {\bibinfo {title} {{Dipolar Spin
  Liquid Ending with Quantum Critical Point in a Gd-based Triangular Magnet,
  arXiv:2301.03571}}} (\bibinfo {year} {2023})\BibitemShut {NoStop}%
\end{thebibliography}
\end{document}